\documentclass[12pt,preprint]{aastex}

\usepackage{graphicx}
\usepackage[]{natbib}
\usepackage{amssymb,subfigure}
\bibpunct{(}{)}{;}{a}{}{,}
\newcommand{\el}[1]{\raisebox{0ex}{\scriptsize #1}}
\begin{document}
\title{SIMULATION OF THE FORMATION AND MORPHOLOGY OF ICE MANTLES ON INTERSTELLAR GRAINS}
\shorttitle{Formation and Morphology of Interstellar Ice Mantles}
\shortauthors{H. M. Cuppen \& Eric Herbst}%
\author{H. M. Cuppen\altaffilmark{1,2} and Eric Herbst\altaffilmark{1,3}}
\altaffiltext{1}{Department of Physics, The Ohio State University, Columbus, OH 43210}
\altaffiltext{2}{Leiden Observatory, Leiden University, P.O. Box 9513, 2300 RA  Leiden, The Netherlands}
\altaffiltext{3}{Departments of Astronomy and Chemistry, The Ohio State University, Columbus, OH 43210}
%-------------------------------------------------------------------------

\begin{abstract}
Although still poorly understood, the chemistry that occurs on the surfaces of interstellar dust particles profoundly affects the growth of molecules in the interstellar medium.  An important set of surface reactions produces icy mantles of many monolayers in cold and dense regions.  The monolayers are dominated by water ice, but also contain CO, CO$_{2}$, and occasionally methanol as well as minor constituents. In this paper, the rate of production  of water-ice dominated mantles is  calculated for different physical conditions  of interstellar clouds and for the first time images of the morphology of interstellar ices are presented. For this purpose, the continuous-time random-walk Monte Carlo simulation technique has  been used. The visual extinction, density, and  gas and grain temperatures are varied. It is shown that our stochastic approach can reproduce the important observation that ice mantles only grow in the denser regions. 
\end{abstract}


\section{INTRODUCTION}
In cold and dense regions of interstellar clouds, infrared absorption studies show the existence of mantles of ices \citep{Williams:1992}, of which water ice is observed to be the main constituent \citep{Whittet:1998,Pontoppidan:2004}. The formation and destruction of the interstellar ices is particularly important in regions where new stars and planets are born. During the first stage of star formation, small cores in a cloud assembly collapse under their own gravitational force, resulting in high gas densities and low temperatures. Under such conditions, virtually all gaseous species heavier than hydrogen and helium accrete onto the grains, where they can undergo chemical processes. In protostellar regions, the ices evaporate and change the nature of the gas-phase chemistry \citep{Caselli:1993,Bottinelli:2004}.  

The formation of water ice-dominated mantles is not yet fully understood. Recombination of smaller species on the surface and condensation of complete molecules from the gas phase are the competing processes, but quantitative estimates of their efficiency under interstellar conditions are lacking. Also, the reason for the apparent absence of water ice in more diffuse clouds, which are exposed to ultraviolet and visible photons, is still not clear. From observations of these regions, we know that the ice mantles constitute approximately less than a few monolayers of water, the current detection limit. %\citep{Tielens:2005}. 
A recent attempt to understand the formation of ice layers was made by \cite{Papoular:2005} who concluded that direct accretion of water from the gas is necessary to initiate ice-mantle formation.  His main explanations for the lack of water ice in the diffuse regions are the high grain temperature and low flux in these regions. 

 As \cite{Papoular:2005}, our goal in this paper is to explain the growth of monolayers of ice in dense sources and the absence of this growth in diffuse sources. With the use of a so-called microscopic Monte Carlo approach (see below), we focus on the accretive recombination mechanism using a set of surface reactions undergone by species that either accrete onto a grain surface or are products of other surface reactions that remain on the grain. The reactants are limited to species containing the elements O and H. We also consider photodissociation processes for surface species caused by ultra-violet photons.  In diffuse sources, these photons are mainly those of the external radiation field, whereas in dense sources, the much smaller flux of photons arises indirectly from cosmic ray bombardment of H$_{2}$, which produces ions and electrons.  The electrons can then excite H$_{2}$, leading to fluorescence \citep{Prasad:1983,Ruffle:2001}.  In translucent and dense clouds, we find that ice mantles grow and we follow their morphology in some detail.  To the best of our knowledge, this is an initial attempt to explain the morphology of ice grown via chemical reactions.

\section{MONTE CARLO METHOD}
\label{MC}
The exact details of the continuous-time random-walk (CTRW) Monte Carlo method are already explained in previous papers \citep{Chang:2005,Cuppen:H2}. We therefore  give only a brief summary here.  The initial grain surface is divided up into a square lattice of adsorption sites, each of which corresponds to a local energy minimum.
We start from a bare surface of amorphous carbon with a high degree of roughness. Carbon is used rather than  olivine because of the more severe lack of available desorption data for the adsorbates we must consider on the latter surface material. { Since the major portion of carbonaceous grains is of an amorphous nature rather than crystalline, we treat the grain as being amorphous. This choice also means that the grains do not have a metallic character, as they would had they consisted of graphite.}  Figure \ref{surf} shows the topology of the surface of the bare carbonaceous grain we use throughout the paper. It is the same as surface (d) in \cite{Cuppen:H2} and surface III in \cite{Cuppen:heating}. \cite{Cuppen:H2} explain the procedure used to obtain this surface.  The roughness is caused by a variety of surface imperfections including vacancies (holes in the surface), edges, islands, and ad-atoms, all of which are shown in Figure 2 of \cite{Cuppen:H2}. The species are confined to a square lattice structure even though we intend to simulate the growth of an amorphous ice mantle. An off-lattice simulation method, where the species are not confined to regular lattice positions,  would seem to be a better choice, but this procedure is computationally expensive. Since we do not have a potential or force field to reliably describe the system, it would not result in much more realistic simulations. The surface plotted in Figure \ref{surf} has $50 \times 50$ lattice sites. Because this size is above the limit where finite size effects play a role  \citep{Chang:2005}, scaling the results to larger sizes does not cause a problem.  Indeed, we scale the results to a standard grain with a radius of 0.1 $\mu$m, which corresponds to a 250-500 times larger grid, depending on the site density chosen. The smaller grid was used to keep the cpu time for the simulations  reasonable.  

Atoms of a species labeled A are deposited on these grains during our simulations with a deposition rate in monolayers (ML) s$^{-1}$ of  \citep{Biham:2001}
\begin{equation}
R_{\rm dep} = \frac{v_A n_{\el{A}}}{4\rho},
\end{equation}
where \(n_{\el{A}}\) is the absolute gas abundance of species A, \(\rho\) is the surface site density, and \(v\) is the mean velocity of the atoms and molecules in the gas:
\begin{equation}
v_A = \sqrt{\frac{8 kT_{gas}}{\pi m_A}},
\end{equation}
with \(T_{gas}\) the temperature of the gas and $m_{A}$, the atomic mass. We use \(\rho = 5 \times 10^{14}\) cm\(^{-2}\) in diffuse clouds and \(\rho = 1 \times 10^{15}\) cm\(^{-2}\) in dense clouds. The latter value is obtained from measurements of high density ice \citep{Jenniskens:1995}. The  value used for diffuse clouds was determined for amorphous carbon \citep{Biham:2001}.   In our simulation, atoms hit the surface under a randomly determined angle. The atom will stick to some environment on the rough surface depending on its incoming angle.  Once on the surface, the atom  can hop over the surface, or desorb. The rates (s$^{-1}$) of hopping and desorption are respectively
\begin{equation}
R_{hop} = \nu_b \exp\left(-\frac{E_{b}}{T}\right)
\end{equation}
and
\begin{equation}
R_{eva} = \nu_D\exp\left(-\frac{E_{D}}{T}\right)
\end{equation}
where \(T\) is the surface temperature of the grain, \(E_b\) and \(E_D\) are the hopping barrier and desorption energy (K), and $\nu_b$ and $\nu_D$ are the attempt frequencies for hopping and desorption, respectively. The frequencies $\nu_b$ and $\nu_D$ are both taken to be $10^{12}$ s$^{-1}$, which is a standard value for physisorbed species \citep{Biham:2001}.  

\subsection{ Binding, Hopping, and Activation Energies}

We assume that species close to step edges or other protrusions of the surface are more strongly bound because of the extra lateral physisorption "bonds" with the surface. Only lateral interactions with the water ice and the bare carbonaceous grain are considered. 

For the binding energy of a particular adsorbate species, labeled A, we use the equation
\begin{equation}
E_{D}^{\el{A}}(i_{c},i_{\el{H}_2\el{O}}) = E^{\el{A}}_s + \alpha i_{c} E_{c}^{\el{A}} + \alpha i_{\el{H}_2\el{O}} E_{\el{H}_2\el{O}}^{\el{A}},
\label{E_D}
\end{equation}
where $E^{\el{A}}_s$ is the  binding energy of A with the particular surface lying beneath it, \(E_{c}^{\el{A}}\) is the binding energy of A with amorphous carbon, \(E_{\el{H}_2\el{O}}^{\el{A}}\) is the binding energy of A with water-ice, $\alpha$ is the fraction of the binding energy that characterizes lateral binding, and 
$i_{s}$  ($s = c, H_{2}O$) is the number of horizontal neighbors of the particle with carbon or ice.  For the hopping energies, we use the equation:
\begin{equation}
E_{b}^{\el{A}}(i_{c},i_{\el{H}_2\el{O}}) = 0.5 E^{\el{A}} + \alpha i_{c} E_{c}^{\el{A}} + \alpha i_{\el{H}_2\el{O}} E_{\el{H}_2\el{O}}^{\el{A}},
\label{E_b}
\end{equation}
which is similar to equation (\ref{E_D}) except for the factor 0.5 in the first term.
Very little is known about the relation between the hopping barrier and the desorption energy. For atomic hydrogen on amorphous carbon and polycrystalline silicate surfaces a factor between the hopping barrier and desorption energy of 0.78 was found \citep{Katz:1999}. This value was obtained by fitting rate equations to temperature-programmed-desorption (TPD) experiments. In most astrochemical models, a much lower value of 0.3 is assumed \citep{Tielens:1987}. Experiments of water-on-water diffusion indicate a similar value to 0.3 whereas CO on water-ice experiments suggest a slightly higher ratio \citep{Collings:2003}. Our value of 0.5 lies closer to this last study and corresponds to the value found by \cite{Ulbricht:2002} for oxygen adsorption on carbon nanotubes. The higher value found by \cite{Katz:1999} could be the result of an average over different hopping barriers. Because our method of lateral bonds already accounts for this effect, we take a lower value. Appendix \ref{dif} discusses the effect of this hopping factor on the results. Note that the lateral bonds must be broken in a hopping movement and that we do not account for the bonds that are re-formed since we assume the barrier  to be independent of these bonds. We do not consider a possible vertical interaction upwards for embedded layers. If a particle has ``upper neighbors'' we do not allow it to evaporate but it can hop, leaving a vacancy, or ``pore'', in the material.

The $\alpha$ factor is taken to be 0.2 in all simulations. Once again, very little is known about what this factor should be for physisorption. We choose our relatively low value to be conservative; non-zero values are needed to explain the high efficiency of H$_{2}$ formation at relevant surface temperatures in diffuse clouds but high values can lead to  complex TPD spectra in the laboratory with multiple peaks \citep{Cuppen:H2}. The factor has the largest effect on the hopping and desorption of hydrogen and on the hopping of oxygen atoms. All other species have such high energies for hopping and desorption that they remain in their sites independently of any changes in energies due to a different $\alpha$ factor.

If two particles encounter each other due to hopping and land in the same lattice site, a reaction can occur. Table \ref{sur_re} lists the surface reactions used to study ice formation,  their estimated activation energies, and a factor $\beta$ that stands for the fraction of products remaining on the surface after a reaction. 
All of the surface reactions  in Table \ref{sur_re} are exothermic. The excess energy that is released in an association reaction (a process with one product) can cause the product to desorb from the surface.  To the best of our knowledge, $\beta$ has been determined experimentally only for the formation of H$_{2}$ on olivine (0.33) and amorphous carbon (0.413) \citep{Katz:1999}, but not on ice. The $\beta$ value for the formation of water from H and OH can be extracted from molecular dynamics simulations of the photodissociation of solid water performed by \cite{Kroes:2006}. They find that a fraction of the H + OH photodissociation products reacts back to water, 0.9 \% of which is released into the gas phase. We take this value of $\beta = 0.991$ for all single-product association reactions. Only products in the outer layers,  which are directly in contact with the gas,  are allowed to evaporate.  A value of $\beta$ near unity is also found in the theory of \cite{Garrod:2006,Garrod:2007}. This fraction is mainly important to remove water from the surface in diffuse clouds. It is of less importance for the H$_2$ production, because H$_{2}$ can evaporate thermally. Since the binding of molecular hydrogen to amorphous ices is similar to its binding to amorphous carbon, the $\beta$-value is expected to be similar on this substrate. We will use the value of $\beta = 0.413$ for this reaction. A higher value will result in a slightly lower hydrogen recombination efficiency \citep{Cuppen:heating}. 

In contrast with might be expected from the case of gas-phase reactions, surface reactions with activation barriers can have noticeable contributions in our treatment, because the number of reaction trials the species can get before one of them hops away multiplied by the probability of reaction can reach unity  \citep{Awad:2005}.  In our approach, if two species land in the same box and can react with activation energy, we do not follow the details of the competition but determine immediately which process  - reaction or hopping - occurs by comparison of the two rates.  There is experimental and/or theoretical evidence that these reactions do indeed occur on cold surfaces \citep{Kroes:2006,Hiraoka:1994,Tielens:1982, Watanabe:2002}.  Here we assume that  the reaction barrier can be treated as a rectangular potential barrier, through which tunneling occurs with a rate (s$^{-1}$)
\begin{equation}
R_{react}=\nu_r P_{react} = \nu_r\exp(-\frac{2a}{h}\sqrt(2\mu E_a))
\label{Preact}
\end{equation}
where $a$ is the width of the barrier, $\mu$ the reduced mass, $\nu_r$ the attempt frequency for reaction, and $E_a$ the reaction activation energy.  We assume that the trial frequency for hopping, desorption and reaction is the same.  We take here a width of 1 \AA. The width of the diffusion barrier is probably larger (D. Woon, private communication) and diffusion therefore most likely occurs through thermal hopping, as in the case of H$_{2}$ formation \citep{Katz:1999}.  The difference in treatment between diffusion and hopping has a large influence on the temperature dependence of reactions with a barrier. The probability $P_{react}$ is temperature independent according to equation (\ref{Preact}) whereas the thermal hopping rate, which is inversely proportional to the residence time, increases with increasing temperature. Tunneling leading to reaction therefore becomes more important as the temperature is lowered, but the decreased hopping rate makes it harder for two reactants to find each other. Detailed experiments and calculations need to be carried out to give more information about how to treat this type of surface reaction.

\subsection{Long-Range Diffusion and Photodissociation}
 If a species tries to hop to a neighboring site that is already occupied and the two species do not react, we allow the particle X to diffuse to the next available free lattice site in the same layer with a probability $P$ given by the empirical equation
\begin{equation}
P = \exp\left(-\frac{E_D(\el{X on H}_2\el{O})}{200}\Delta x\right),
\label{P}
\end{equation}
where $\Delta x$ is the distance in sites to the next available free site, and $E_{D}$ is the binding energy in K.  This equation is chosen so that small, weakly-bound species such as atomic and molecular hydrogen can diffuse through porous ice structures, with a probability that  rapidly decreases with increasing distance. \cite{Andersson:2006} showed in molecular dynamics simulations that atomic hydrogen can move through the ice matrix after photodissociation of water, when the hydrogen atom is translationally excited. Eq.~\ref{P} allows this motion to occur. The probability is an estimate based on the arguments mentioned above, since information for thermalized hydrogen at in the range of 10 to 20 Kelvin is lacking. Eq.~\ref{P} does not contain an attempt frequency  because this probability is only called if a normal hopping event cannot occur since the neighboring site is already occupied. The attempt frequency is therefore already in the hopping rate as is the temperature dependence. The particle can diffuse in all six directions. If the particle does not diffuse, it remains in its original position. Using binding energies discussed in the next section,  the probability of diffusion of H over two sites $P$ is $4.8\times 10^{-3}$ while it is  $1.5\times 10^{-8}$ for the two-site diffusion of a heavier ozone molecule.

Surface species can be destroyed by photodissociation, either directly or by cosmic-ray-induced photons.  We assume that the photodissociation rate of the surface species is the same as in the gas phase, and is given by the standard expressions
\begin{equation}
R= \alpha_{photo}\exp(-\gamma_{photo}A_V)
\end{equation}
and
\begin{equation}
R= \alpha_{cr}\zeta
\end{equation}
for direct and  cosmic-ray-induced photodissociation, respectively. The constants are given in Table \ref{dis_re} and for the cosmic ray ionization rate $\zeta$ we use $1.3\times 10^{-17}$ s$^{-1}$. These assumptions are only partially in agreement with molecular dynamics calculations for ice done by \cite{Kroes:2006}, who show that the photodissociation cross section for water ice does not have exactly the same dependence on photon energy as is the case for gas-phase water. In our simulation, all molecules in the grain mantle can be photodissociated and a shielding mechanism is not included. \cite{Nguyen:2002} assumed a limit of 100 monolayers that could be reached by photodissociation. In our simulations, only in dense regions where photodissociation plays a minor role can these thicknesses be reached. 

We further assume that the heavy product fragments  remain on the surface after photodissociation.   These product fragments can, however, react  leading to the possibility of desorption in association reactions.  Following \cite{Kroes:2006} we assume that 40 \% of the atomic hydrogen formed by photodissociation of H$_2$O evaporates. This value is an average of the value they obtained for the first three layers. We also include their indirect mechanism for the photodesorption of water  in which a thermally excited hydrogen atom kicks out one of the water molecules with an overall probability of  0.04\%.

\section{ENERGIES}
\label{energies}
Table \ref{E_eva} contains the binding (desorption) energies used between assorted adsorbates and surfaces  in our calculations.  
Many interaction energies are needed to describe the system, not just between the different species with the bare carbonaceous substrate but also with ice and with the other adsorbed species. The problem is, however, that many of these interaction energies are poorly known.  For amorphous carbon, as far as we know, only the desorption energies for H and H$_2$ are experimentally determined \citep{Pirronello:1999,Katz:1999}, with values of 658 K for H and 542 K for H$_{2}$.  { Additional studies have been undertaken with graphite and carbon nanotubes. Even though our model grain consists of amorphous carbon, we occasionally have to rely on  studies based on other forms of carbon. }
\cite{Ghio:1980} calculated the van der Waals interaction between graphite and atomic hydrogen and found an energy of 503 \(\pm\) 6 K, which is in excellent agreement with the overview of experimental data given by \cite{Vidali:1991}, in which the best estimate is given as 499 \(\pm\) 3 K.  For the H$_2$ interactions with carbonaceous material we found two additional studies. \cite{Han:2004} used DFT methods to calculate a value of 400 K for H$_2$ on carbon nanotubes, while the survey of \cite{Vidali:1991} gives a best estimate of 660 \(\pm\) 6 K for H$_2$-graphite.  In both cases, we use the amorphous carbon results. 

For the other species, there are no amorphous carbon data and we have to make use of theoretical or experimental results  for graphite, carbon nanotubes, or other forms of carbon. \cite{Papoular:2005} performed chemical simulations to obtain the interactions between O, OH and H$_2$O and typical hydrocarbon surface functional groups. He found energies of 300 K and 500 K for O and OH, respectively, while for water he found a range of energies between 250 and 1000 K.  The O and OH values seem rather low compared with the energies for the much lighter species H and H$_{2}$. The calculations were performed between small single molecules and the absorbate. In reality, there will probably be many more contributions from other atoms in the substrate resulting in a higher interaction. We prefer to use the estimates of  
 800 K \citep{Tielens:1982} for O and 1360 K \citep{Allen:1975} for OH.  For H$_{2}$O, there is another theoretical value by \cite{Picaud:2004},  who determined the energy to be 800 K.  
Experimental results give 1870 K \citep{Avgul:1970} and 2150 K \citep{Gale:1964} for the same system. We use 2000 K based on the experiments.   For O$_2$, we use the value of 1440 K from \cite{Ulbricht:2002} obtained by TPD experiments on graphite, while for the species O$_{3}$, O$_{2}$H, and H$_{2}$O$_{2}$, we simply add the binding energies of their constituent parts:
\begin{eqnarray}
E_D(\el{O}_3) &= &E_D(\el{O}_2)+ E_D(\el{O}),\\
E_D(\el{O}_2\el{H}) &=& E_D(\el{OH})+ E_D(\el{O},)\\
E_D(\el{H}_2\el{O}_2) &= &E_D(\el{O}_2\el{H})+ E_D(\el{H}).
\end{eqnarray}

Let us next consider the energies of the species on solid water/ice. The problem here is that there are many types of ice, both in structure and in morphology and porosity, which makes it difficult to compare different experimental results. The porosity has two effects: (i) the particles in the pores cannot desorb directly but will perform a random walk on the wall of the pores until they reach the outer surface, and (ii) the increased roughness in the pores means that  the number of  binding sites with high energy is probably larger. Since we already account for the latter effect by considering multiple types of sites, we use non-porous results as much as possible.  A considerable amount of data is available for molecular hydrogen. \cite{Hornekaer:2005} and \cite{Dulieu:2005} find a large distribution of energies using TPD experiments. If we only consider the mean value for the non-porous substrate we get 440 K from \cite{Hornekaer:2005} for  D$_2$ and 520 K and 550 K from \cite{Dulieu:2005} for H$_2$ and D$_2$, respectively. These values are close to the theoretical value obtained by \cite{Hollenbach:1970} of 550 K and the experimental value of \cite{Sandford:1993b} of 550 $\pm$ 35 K, based on a mixture of water and methanol ice.  The results of \cite{Perets:2005}, however, are in disagreement with this cluster of values.  Performing TPD experiments for HD and D$_2$ on high density ice (HDI) and low density ice (LDI), they found  three types of sites for the LDI, which has a higher porosity and one type of site for the HDI with binding energies  798 K (HD) and 836 K (D$_2$).  We use 440 K, the lower limit of the distribution and the peak energy found by \cite{Hornekaer:2005}. Since we include higher energy sites near step edges in our model, we still obtain a distribution of desorption energies in this way. 

For H on ice, we found three theoretical values:  \cite{Hollenbach:1970}, \cite{Buch:1991} and \cite{Al-Halabi:2002} determined energies of  450 K, \(\approx\) 500 K and 400 $\pm$ 50 K, respectively. {The values of \cite{Hollenbach:1970} and \cite{Al-Halabi:2002} are for crystalline ice whereas the \cite{Buch:1991} value is for amorphous ice.  \cite{Perets:2005} determined a significantly higher value, 720 K, from TPD experiments on HDI.   We  use 650 K, which is a new value calculated by \cite{Al-Halabi:pub} for the binding energy of atomic hydrogen on amorphous ice. They found that the binding on amorphous ice is significantly higher than on crystalline ice due to the surface structure. Even a surface without protrusions of amorphous ice has several small cavities that will bind the hydrogen more strongly whereas the crystalline ice is even flat on an atomic scale.  

For water on amorphous ice we use 5640 K, which is the result obtained by \cite{Speedy:1996}, and is in excellent agreement with a value of \(\approx\) 5600 K found by \cite{Fraser:2001}. For O$_2$ we use 1000 K, which is an average value obtained from the TPD data by \cite{Ayotte:2001} and \cite{Collings:2004}. For O and OH we again take the estimates from \cite{Tielens:1982} and \cite{Allen:1975}, respectively, and for O$_{3}$, O$_{2}$H, and H$_{2}$O$_{2}$, we again add the binding energies of their constituent parts. 

Let us now consider H$_{2}$ as a substrate, since, as we will see, its abundance in the mantles can be considerable. \cite{Vidali:1991} give an estimate of the desorption energy for H from H$_2$ of 35.4 $\pm$ 5.8 K + $kT$ which leads to 45.4 K at 10 K. The rest of the desorption energies from H$_{2}$ are, starting from the ice desorption energies, scaled according to the equation:
\begin{equation}
E_D(\el{X on H}_2)=\frac{E_D(\el{X on H}_2\el{O})}{E_D(\el{H on H}_2\el{O})}E_D(\el{H on H}_2).
\label{ED X}
\end{equation}                                 
Once again, energies involving  O$_3$, O$_2$H and H$_2$O$_2$ are formed from their fragments.

{ The reason the desorption energy of  O$_2$H is considered to be the sum of the desorption energies of OH and O instead of O$_2$ and H, which are better known, is the capability of O$_2$H to form H-type bonds. We are aware that  desorption energies obtained in this manner have a large uncertainty. Fortunately, our simulations show the uncertainties in the desorption energies of the strongly bound species to have very little effect on the final result, since these species will remain on the surface at these temperatures within the time-scale of the simulation regardless of their exact desorption energies. }

\section{RESULTS}

 In all simulations the water was formed through surface reactions and not by direct accretion from the gas phase, even though gas-phase water can be detected in some dense objects.  This exclusion is easily justified. A typical value obtained in gas-phase models for the fractional abundance of water is at most 10$^{-6}$.  For a cloud of density 10$^{4}$ cm$^{-3}$, and a  standard site density, we then calculate that the flux rate for gaseous water at 10 K upon a grain is 3 $\times 10^{-14}$ ML s$^{-1}$, so that it takes roughly 10$^{6}$ yr to deposit a single monolayer and the enormous time of 10$^{8}$ yr to deposit the 100 monolayers typically associated with dense cloud mantles. The time scales are much longer for diffuse clouds.  The flux of oxygen atoms arriving on grains is much higher than the flux of water molecules, typically by a factor of 100 in dense clouds.  With a high efficiency of conversion to water by surface reactions, 100  monolayers of water can then be produced in 10$^{6}$ yr, in reasonable agreement with our dense-cloud results discussed below.  So, achieving significant mantle abundances in reasonable times does require surface reactions.

\subsection{Diffuse and Translucent Clouds}

We have performed a set of simulations for different values of the visual extinction $A_{V}$, total hydrogen density ($n_{\rm H}$), and gas and grain temperatures. Figure \ref{res} plots the surface abundances in monolayers for the different species in our model as functions of time. In this figure, the results for diffuse and translucent clouds are contained. All panels are labeled with a letter followed by a number. The panels starting with the same letter are grouped in rows and have the same visual extinction, which increases from top to bottom. The digits along a row indicate the different physical conditions. For the three leftward panels, the density is fixed but the temperatures decrease as one goes from left to right, while the rightmost panel  (e.g. A4) has the same temperature as the second of the three left panels (e.g. A2)  but a higher density. The exact physical conditions for each of the simulations are given in Table \ref{phys}. In all cases shown in Figure \ref{res}, we assume the cloud gas to be initially atomic, with an oxygen abundance of $3.0 \times 10^{-4} n_H$ that of hydrogen \citep{Nguyen:2002}. The calculations are run for sufficiently short times that the gas-phase abundances remain unchanged. In all cases we start with a bare carbonaceous grain. 

Figure \ref{res} clearly shows that in diffuse regions ($A_{V} =0.5, 1$, rows A and B) less than 1 ML of water-ice is present at a time of $3 \times 10^{5}$ yr, the largest time studied.   For panels A1-A3, the water-ice abundance has reached steady-state conditions while in panel A4, which refers to a higher density, steady state seems to be reached around the end of the simulation. In panels B1 and B2 the water layer is still growing, albeit slowly, while in panels B3 and B4 the water abundance is decreasing. Although water-ice is not destroyed chemically in our model, a steady-state results when destruction by photons and desorption from the mantle balance formation. Water-ice is the most abundant species for the diffuse clouds in rows A and B; nonetheless, other species can be seen in the panels.  Despite its low desorption energy, molecular hydrogen can reach mantle abundances of a significant fraction of a monolayer in diffuse clouds.  
In the translucent clouds, rows C and D, the water-ice abundance exceeds 1 ML by the end of $3 \times 10^{5}$ yr and is still growing. Water-ice is dominant in all panels except D4, where OH has a comparable total abundance.  As far as we know, no data from observations is available about the abundance of OH present on grains under comparable conditions.

 If we look at the influence of the physical conditions on the formation of ice for diffuse and translucent clouds, it can be concluded that  the gas and grain temperature affect the ice formation mechanism only to a small extent. In particular,  a low surface temperature results in a slightly thicker ice mantle.  In general, the ice thickness is mostly determined by photodissociation under the influence of the radiation field and by the gas density.  Thus, as one goes down a column  and to the rightmost panel of any row (higher density), the water-ice abundance increases.  
Figure \ref{res} shows a clear difference in the composition of the ice layer depending on the visual extinction. At $A_{V} =0.5$ (A panels) the mantle mainly consists of water, and molecular hydrogen. For increasing $A_{V}$ the amounts of OH and O$_3$ increase. The surface abundance of atomic oxygen appears to peak at $A_{V} =2$ (C panels). 

Fig.~\ref{images} shows a vertical slice of the ice at $3 \times 10^{5}$ yr for four different conditions -- panels A2, B2, C2, and D2 -- in which visual extinction and gas density increase.  The different molecules are color coded in the panels, while gray represents the carbon substrate and black represents the void. One sees immediately that the mantle for A2 is virtually non-existent, while that for D2 has grown to a significant number of monolayers. The mantles are certainly porous; there is a lot of black interspersed among the color-coded molecules. Due to the strong radiation in all simulations, the ice goes through many cycles of formation and destruction.  Let us take a more detailed look at mantles C2 and D2.  The top layers of the mantles are primarily water (white color).  As one goes deeper into the mantle, one notices more green (OH), and finally towards the bottom one finds lots of magenta (ozone).  In understanding this dependence of abundance on mantle height, the topology of the mantle is important.
 Blocking of the evaporation of the hydroxyl radicals and oxygen atoms after photodissociation and obstruction of the deposition of fresh hydrogen from the gas phase are the main reasons that allow the large abundances of OH and O$_3$ to be formed away from the surface. Atomic hydrogen leaves the surface either by direct desorption after photodissociation or in the form of H$_2$ that can diffuse through the pores. We used the same desorption probability of H after water photodissociation of 0.4 throughout the ice. If H atoms accreting from the gas have a greater penetration depth or if a depth dependence of the desorption probability upon photodissociation was considered, the large abundance variance with depth we predict could be partially reduced.  Unfortunately, very little experimental information on the structure of water ices formed through surface reactions under interstellar conditions is available so that our images of the ice mantles cannot be easily compared with real ices. 
 
The common gas-grain codes \citep{Ruffle:2001} do not keep track of the molecular structure of the ice layers so structural variations cannot be found. The mean field approach of most of these methods allows all species to react with each other independently of their position.  The mantle abundances of the hydroxyl radical and ozone will therefore be lower in these simulations using the same surface reaction network since their reactions with H atoms will occur more readily.  Even in our variegated ices, the mantle abundances of OH and O$_3$  may decrease to some extent if other reactive species like CO are included and are allowed to react with OH and O$_3$ in our reaction network.  For the latter case these reactive species must be able to diffuse in the lower layers of the ice.

Molecular hydrogen formation from the recombination of two hydrogen atoms on a surface is the most important reaction in the diffuse interstellar medium. The recombination efficiency for H$_{2}$ formation is defined by the ratio of twice the hydrogen molecules formed and leaving the surface divided by the flux of hydrogen atoms.  This efficiency has been determined over the time interval that steady state is achieved for the 16 sets of diffuse and translucent physical conditions (A-D) given in Table \ref{phys} and the results are given in Table \ref{etha} by the left value in each column. The values on the right indicate the efficiency for the same densities and temperatures but without oxygen in the system, which means without ice being formed. There is a clear difference in efficiency, with the ice surface being less efficient in most cases. This is due to two effects: the first is that the hydrogen atoms are also used to form water, but this is only of minor importance. The larger effect is caused by  the changing energetics  between the water layers and the carbonaceous substrate. In particular, the binding energy of H on an ice surface is slightly less than on a carbonaceous grain and also the surface roughness changes due to the build-up of the water layers.   In these calculations,  we only use an $\alpha$ of 0.2 for the lateral bond strength. A higher $\alpha$ would result in a more efficient molecular hydrogen production because it causes the hydrogen atoms to be more strongly bound and to reside longer on the surface \citep{Cuppen:H2}.   The efficiencies for lower surface temperatures (12-15 K) are generally still high enough to produce a reasonable amount of H$_{2}$  \citep{Cuppen:heating} but the efficiencies are clearly degraded at higher temperatures.   For smaller grains, we expect stochastic heating to reduce the formation of ice since both oxygen and hydrogen atoms will usually evaporate when a photon hits a grain. This effect would cause  the recombination efficiency for molecular hydrogen to approach those  reported in  
\cite{Cuppen:heating}.

\subsection{Dense Molecular Clouds}
Figure \ref{des} is similar to  Figure \ref{res} but shows results for higher visual extinctions ($A_{\rm V}$  = 5, 10) and densities ($ 5.0 \times 10^{3}$ cm$^{-3}$ -- $5.0 \times 10^{4}$ cm$^{-3}$). The maximum time considered is $1 \times 10^{5}$ yr, at which time  steady-state abundances have not yet been reached for the major constituents.  Again the visual extinction is varied vertically while temperature and density for a fixed extinction are varied horizontally. The exact conditions are once again given in Table \ref{phys}.  We assume that the hydrogen in the gas phase is mostly molecular.  A relative abundance of $1\times10^{-4} n_H$ for atomic H is used.  Although it is assumed that the mantles grow upon a carbonaceous substrate, it is more likely that they grow upon ice mantles of the type discussed in the previous section for translucent clouds.  It can be seen in Figure \ref{des} that in seven out of the eight panels, the growth of water-ice dominates the mantles, while in one panel (F4) the abundance of H$_{2}$ overtakes that of water at later times.  Mantles of up to 75-80 monolayers are developed.  

In the panels, a general tendency can be observed that a lower temperature (compare E1, E2, and E3) and a lower density (compare E2 and E4) result in less ice. This is in contrast with the results in Figure \ref{res}, where the amount of ice increases for decreasing surface temperature. In the diffuse clouds the grain temperature is in the regime where H atoms easily evaporate. Reducing the surface temperature will therefore result in more hydrogen on the surface increasing the efficiency of the hydrogenation reactions. The lower grain temperature in the dense clouds lies in the regime where the desorption of  hydrogen molecules starts to slow down and these molecules occupy a large part of the mantle, slowing down the rate of hydrogenation reactions. Although laboratory simulations of interstellar ice reactions show that hydrogenation reactions are still possible in the presence of H$_2$  because  the hydrogen can still penetrate H$_2$ layers \citep{Fuchs:prep},  the reactions are slowed down, which results in an optimum temperature for the reaction rate. For CO + H, this optimum lies between 12 and 15 K \citep{Fuchs:prep}.

Since the atomic hydrogen abundance is much lower than the molecular hydrogen abundance, the main route to form water ice changes in dense clouds. Table \ref{reactions} gives the contributions of the three reactions that form water: H + OH, H + H$_{2}$O$_{2}$ and OH + H$_{2}$.  This table clearly indicates that the main route for formation is through the association of H and OH in the diffuse and translucent clouds. Since H is abundant in the gas here and since this reaction is barrierless, this mechanism is the most apparent one. However in dense clouds, the reaction between H$_2$ and OH becomes the most common route and also the reaction of H with H$_2$O$_2$  makes a significant contribution. It seems surprising that in a dense molecular cloud a reaction with an activation barrier is so important for the formation of water ice. This is different from the gas-grain network simulations where the barrierless reaction remains the main route \citep{Ruffle:2001} along with some other surface reactions that were not included in this work. The difference is mainly due to the difference in treatment of the activation barriers, as explained earlier: because of our approach, in which the competition with diffusion is modeled directly, reactions with a barrier can have enhanced rates.  In the approach used in the models, tunneling under the activation energy barrier is simply treated as a factor in the expression for the rate coefficient.

Figure \ref{imagesDense} gives two vertical snapshots of ice mantles. Again the surface structure is very rough and porous.  Unlike the diffuse cloud ices, however, the structures demonstrate a skyscraper, or protrusion, effect, which is even more pronounced for lower diffusion rates, as discussed in Appendix \ref{dif}.  Since the protrusions contain a significant amount of molecular hydrogen and since we have not seen them up to now, it is likely that they can only be associated with cold dense cloud conditions with the diffusion to desorption of 0.5 as is used here. The width and height of the protrusions are associated with the  random angle deposition method.   With this method, the protrusions continue to grow since they prevent incoming atoms from reaching the surface via a shadowing effect.  It is likely that the tendency to form  such noticeable skyscraper structures will be mitigated in the interstellar medium for several reasons.  First, the grain surfaces are curved, which reduces the shadowing effect. A standard grain with a radius of 0.1 $\mu$m has a circumference of approximately 2000 sites, which means the snapshots in  Figure \ref{imagesDense} represent arcs of 9$^\circ$. For a small grain of $r = 0.02 \mu$m, it corresponds to 45$^\circ$. Secondly, the rate of diffusion considered in our models is likely a lower limit because diffusion also occurs in the bulk structures that are formed. These will result in a smoother surface with less molecular hydrogen (see the following paragraph).  To confirm our hypothesis concerning the random angle deposition method, we also performed simulations with perpendicular deposition. Large single protrusions containing a high percentage of H$_2$ were not obtained under these conditions; only structures similar to the left panel in Fig.~\ref{imagesDense} were found. The sticking fraction (fraction remaining in the mantles) in the simulation with the perpendicular deposition is much higher and a thicker layer of ice is formed. Angular deposition gives more irregular structures at the surface that have a weaker binding energy and can therefore desorb more easily, resulting in a lower sticking fraction.

Our results can profitably be compared with another simulation.  
\cite{Kimmel:2001II} studied the porosity of ice mantles formed by deposition of water molecules using a ballistic algorithm. In their algorithm they deposited the molecules under specific angles to simulate a molecular beam or with varying angles to simulate a background pressure.  One of their conclusions is that ices deposited from a background pressure  are very rough and porous, in agreement with our result that random angle deposition results in such structures.  
 \cite{Kimmel:2001II} also let a molecule anneal immediately after its deposition. Their annealing algorithm consists of several cycles in which the deposited species hop to energetically more favored sites. Annealing results in denser ices (fewer narrow skyscrapers).  We do not include annealing in our simulations, but we do include diffusion of H and O atoms whereas the water molecules of 
 \cite{Kimmel:2001II} do not diffuse thermally.  In agreement with the observations of  \cite{Kimmel:2001II} is our result that more diffusion results in denser ices, as can be seen by comparing the broad skyscrapers in Figure \ref{imagesDense} with the narrow ones discussed in Appendix \ref{dif}.

\section{DISCUSSION}

Using the CTRW Monte Carlo method, we have shown that mantles of water and other ices can grow efficiently in cold and dense regions of the interstellar medium whereas only a small amount, in the submonolayer regime, is formed under the conditions of diffuse interstellar clouds.  Ices in translucent sources occupy an intermediate regime. We have also looked carefully at the morphology of the ices.  This initial exploration is, however, dependent on a variety of poorly constrained parameters such as barriers to diffusion, which are generally less well understood than adsorption energies. With our standard model, the ices appear to be much denser in diffuse and translucent sources, although there are many pores in the interior.  Also, there is a vertical stratification in these ices, in which water-ice dominates in the uppermost monolayers while unusual species such as OH and O$_{3}$ are abundant in the lower monolayers. In the dense sources, the ices, which are dominated by water and occasionally H$_{2}$, appear to develop a skyscraper-type structure, which is somewhat dependent on the rate of diffusion adopted.  This very rough structure is probably exaggerated somewhat in our calculations because we do not properly include diffusion inside the bulk.  Although the water ice in the diffuse and translucent clouds is produced by the association reaction  between H atoms and OH radicals, which is customarily assumed to be the dominant source of water, the situation in the dense clouds, which contain low abundances of gas-phase hydrogen atoms, is quite different.

Observations in dense cloud assemblies such as Taurus show no water ice below a threshold value of $A_V=3.2$ and linear growth of the column density of water ice above this value \citep{Whittet:2001}. In particular, with a least-squares-fit, the expression
\begin{equation}
N(\el{H}_2\el{O}) = q (A_V^{obs} - A_{th})
\label{N(H2O)}
\end{equation}
can be determined  \citep{Whittet:2001}, where $q = (1.30 \pm 0.04 \times 10^{17})$ cm$^{-2}$ and $A_{th} = 3.2 \pm 0.1$, using the relationship
\begin{equation}
N(\el{H}_2\el{O})  = 1.8\times 10^{18}\rm{[cm^{-2}]} \tau_{3.0}.
\label{N(tau)}
\end{equation} 
Relating these observations to the results of our simulations is not completely straightforward. The measured threshold value of $A_V= 3.2$ is observed through the cloud and gives an edge-to-edge value. The observed ice content is the integrated contribution along a  line-of-sight that typically contains heterogeneous material.  The visual extinction experienced by most material along this line is lower, on average one-half of the edge-to-edge value. 

We make the assumption that the extinction along any line of sight is dominated by one object, defined by the physical conditions used for our models and listed in Table \ref{phys}.   The number of monolayers at 10$^{5}$ yr for the dense objects and $3 \times 10^{5}$ yr for the diffuse and translucent objects is converted into column densities of ice using the gas-to-dust ratio for standard-sized grains and typical cloud depths \citep{Nguyen:2002}.  The results are plotted in Figure \ref{IcevsAv} at the six different visual extinctions used in our  simulations.  The several different results at each  extinction correspond to the different physical conditions chosen.  Also depicted in the figure are two dashed lines that represent the results of equation (\ref{N(H2O)}) assuming that the observed edge-to-edge extinctions correspond to our extinctions -- $A_V^{sim}= A_V^{obs}$  -- and that they correspond to twice our extinctions: $A_V^{sim}= \frac{1}{2}A_V^{obs}$.
 Finally, there is a solid line representing roughly the lower detectable limit for water ice based on the H$_2$O optical depth at 3$\mu$m towards Cygn OB 2 no. 12 of $\tau <0.02$ \citep{Whittet:1997}. This optical depth corresponds
to an $N(\el{H}_2\el{O})$ of $3.6 \times 10^{16}$ cm$^{-2}$, via Eq.~\ref{N(tau)}, which is on the order of a few monolayers.  Assuming that the typical extinction seen by ice along a line-of-sight is indeed 1/2 the edge-to-edge value, our results should approach the upper dashed line shown in Figure \ref{IcevsAv}. The plot indicates, however, that our simulations produce less ice than needed by a factor of $\approx 3-10$.  

This shortfall can be accounted for by increasing the depth of our sources and, for dense clouds, extending the simulations to later times since the calculated ice abundances are still increasing rapidly at our latest time of 10$^{5}$ yr.  However,
cosmic ray desorption starts to become important after 10$^5$ yr and part of the grain mantle can desorb in this way \citep{Herbst:2006}.  It may be that we underestimate the efficiency of water-ice formation, especially at intermediate values of the visual extinction.  If so, we can produce more ice if the rates of diffusion are enhanced, as discussed in Appendix \ref{dif}. Also a decrease of the shadowing effect due to the presence of curvature, as discussed earlier, can result in thicker ice layers on small grains as we saw in the perpendicular deposition simulations.  Yet another possibility is that some ice is indeed accreted from gaseous water, possibly in the cold gas following a shock, which would allow water to be produced in the gas via the reaction between OH and H$_{2}$, despite its activation energy.

E.H. would like to thank the National Science Foundation (US) for supporting his research program in astrochemistry.

\appendix
\section{THE INFLUENCE OF THE HOPPING BARRIERS}
\label{dif}
A ratio of 0.5 between the diffusion barrier and desorption energy has been assumed throughout the paper. This appendix shows the simulation results using an alternative ratio of 0.78, generalizing the result  obtained by \cite{Katz:1999} for atomic hydrogen on olivine. 

The results are shown in Figures \ref{desA} to \ref{imagesDenseA} and in Table \ref{ethaA}.  Figure \ref{desA} shows surface abundances for diffuse and translucent cloud conditions as a function of time, while Figure \ref{resA}   does the same for dense cloud conditions. Figure \ref{imagesA} contains vertical cross sections of ice under diffuse and translucent conditions, while Figure \ref{imagesDenseA} contains such images for dense cloud conditions. Table \ref{ethaA} lists H$_{2}$ recombination efficiencies for diffuse and translucent clouds. As can be seen by comparing these figures and table with the analogous ones for our standard hopping barrier, there are important differences in the amount of ice, the recombination efficiency for molecular hydrogen formation, and in the structure of the ice between these results and the previous ones.  Let us consider them in turn.  

Considering the structure of the mantles, the ice is less dense for the higher hopping barrier, as can be clearly seen by comparing Figures~\ref{images} and \ref{imagesA} and Figures~\ref{imagesDense} and \ref{imagesDenseA}. This can be explained by realizing that for the lower hopping barrier, oxygen atoms can diffuse to more favorable positions (with more neighbors) before reaction.  

Large protrusions as the one shown in the left panel of Figure~\ref{imagesDenseA} can be observed for both diffusion ratios, but they are more common and higher for the higher ratio, and are even present in translucent sources. The main reason is that the mobility of species like O, H, and H$_2$ is limited to such an extent that they cannot move to energetically more favorable position before it is covered by another particle. In cold dense clouds this effect is the strongest since the surface temperature is low and the flux of, especially, molecular hydrogen is very high.
As explained in the main text, the formation of these protrusions is amplified by self-shadowing of the surface by the protrusion. Since a real interstellar grain surface will have a curvature, which becomes significant for smaller grains of approximately 0.02 $\mu$m, this shadowing effect will be less in reality.  

More ice (see Figure \ref{desA} in comparison with Figure \ref{res}) and more molecular hydrogen are formed for the higher ratio. This is consistent with previous results for molecular hydrogen formation. If we compare the right columns in Tables \ref{etha} and \ref{ethaA}, we see that also in the case where only a hydrogen flux is present, the production of molecular hydrogen is much higher.

\section{THE INFLUENCE OF THE ROUGHNESS OF THE GRAIN}
This appendix contains the results of simulations starting with a less rough grain than our standard one. With this smoother surface,  the running time of the simulations becomes significantly longer. This effect occurs because on the smoother surface the average hopping time is smaller so that the simulation program spends more of its time evaluating hopping events. For this reason we performed simulations only for a few of the conditions in Table \ref{phys}. Surface abundances vs time are shown in Figure \ref{islands} for panels A3, B3, C3, and D3.

In general, there is little difference between these results and our standard ones.  The amount of ice that is formed is very similar for all simulations, except for the initial peak around $5 \times 10^4$ yr for B3 when the smoother surface is used. The surface abundances of the other species change slightly in the very diffuse regions. A3 has less H$_2$ and B3 less O for the smoother surface.  In the other two cases, the surface is completely covered with ice, which determines the energetics so that the difference in surface abundances between the smoother and the rougher surfaces becomes negligible.

\begin{figure*}
\plotone{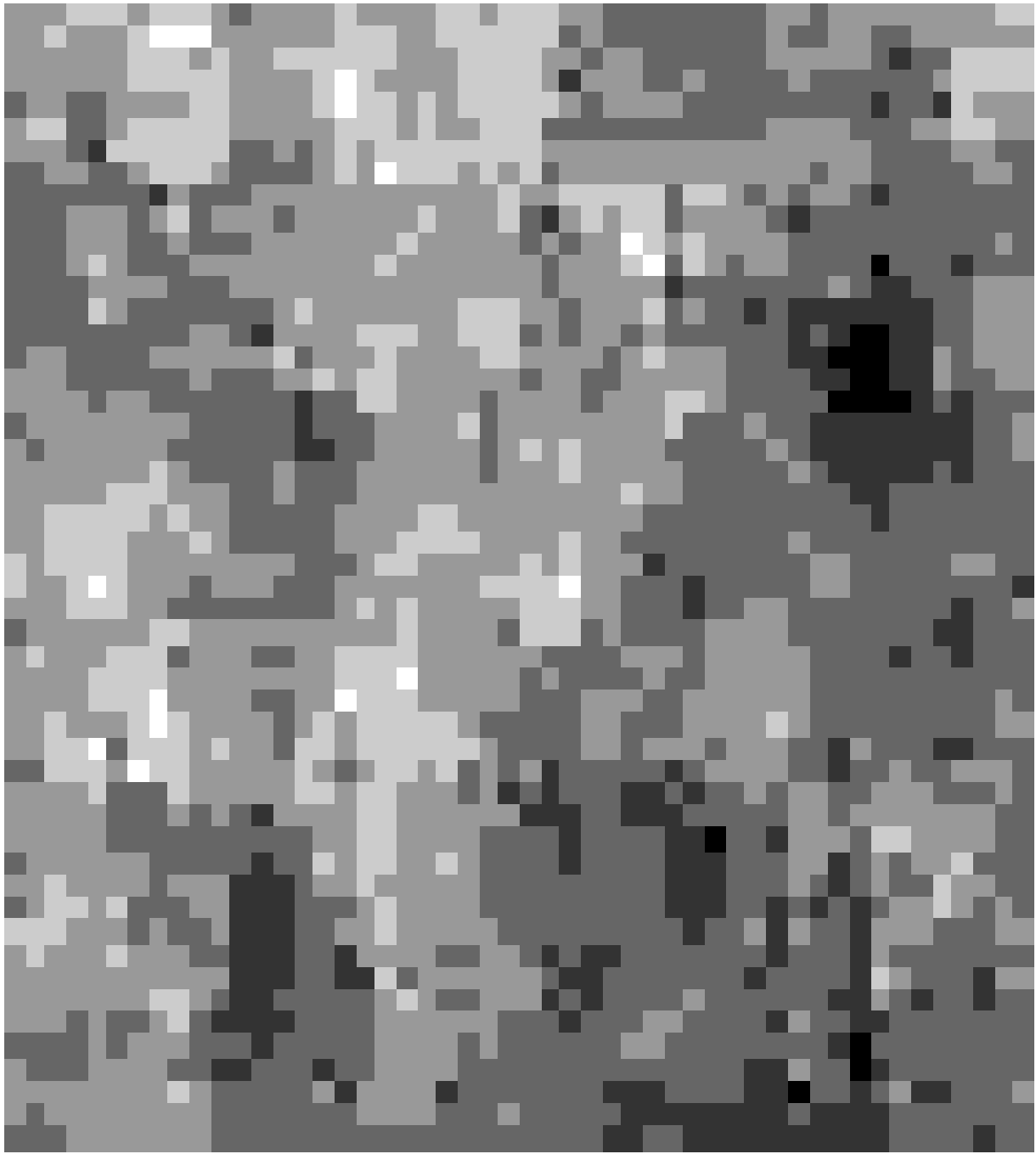}
\caption{The surface topology of the bare carbonaceous grain, represented by a lattice of adsorption sites.  The lighter colors indicate the higher areas. See \cite{Cuppen:H2} for a detailed discussion of how the surface was generated.}
\label{surf}
\end{figure*}

\begin{figure*}
\includegraphics[width=0.8\textwidth]{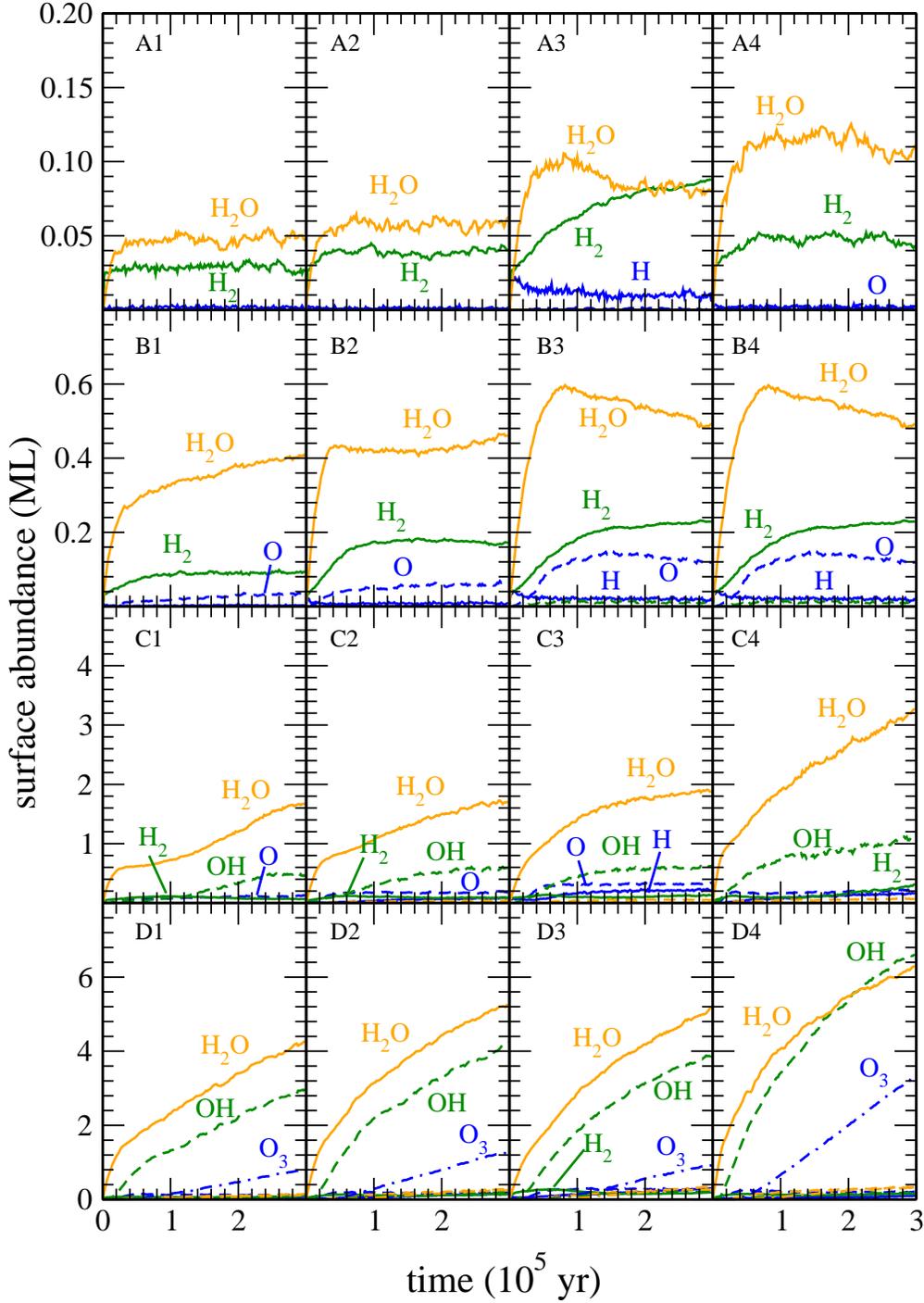}
\caption{The surface abundance as a function of time for different conditions in a diffuse or translucent molecular cloud. See Table \ref{phys} for physical conditions in each panel.}
\label{res}
\end{figure*}

\begin{figure*}
\plotone{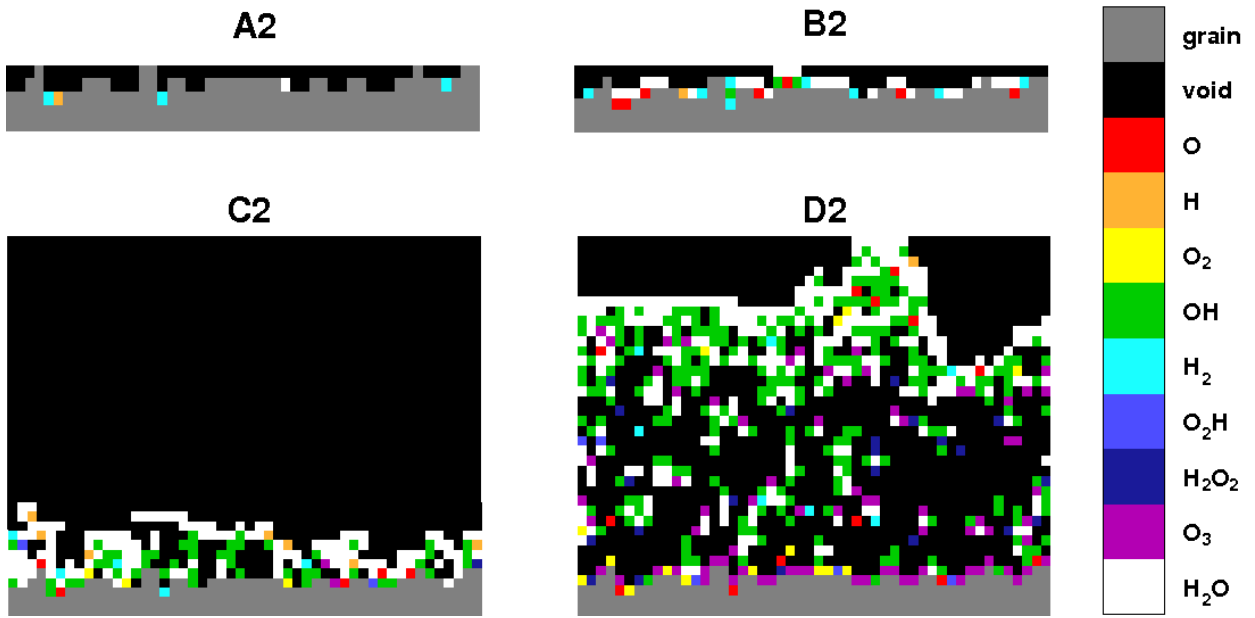}
\caption{Vertical cross sections of the ice mantles after $3 \times 10^5$ yr for four different physical conditions in diffuse and translucent clouds (A2, B2, C2, D2; Table \ref{phys}). The molecules are color coded.}
\label{images}
\end{figure*}

\begin{figure*}
\plotone{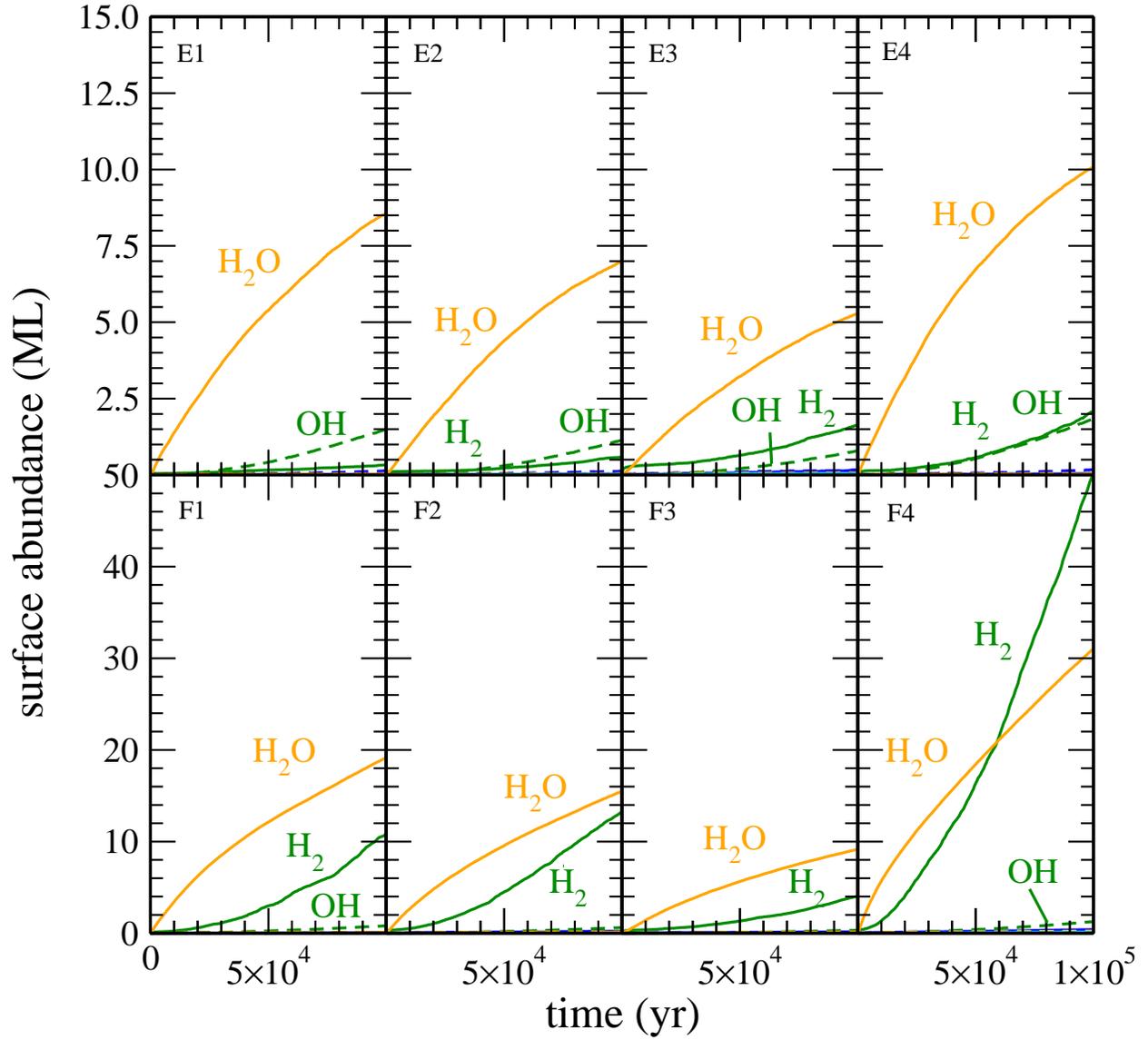}
\caption{The surface abundance as a function of time for different conditions in a dense molecular cloud. The notation a(b) implies $a\times10^{b}$.  See Table \ref{phys} for physical conditions in each panel.}
\label{des}
\end{figure*}

\begin{figure*}
\plotone{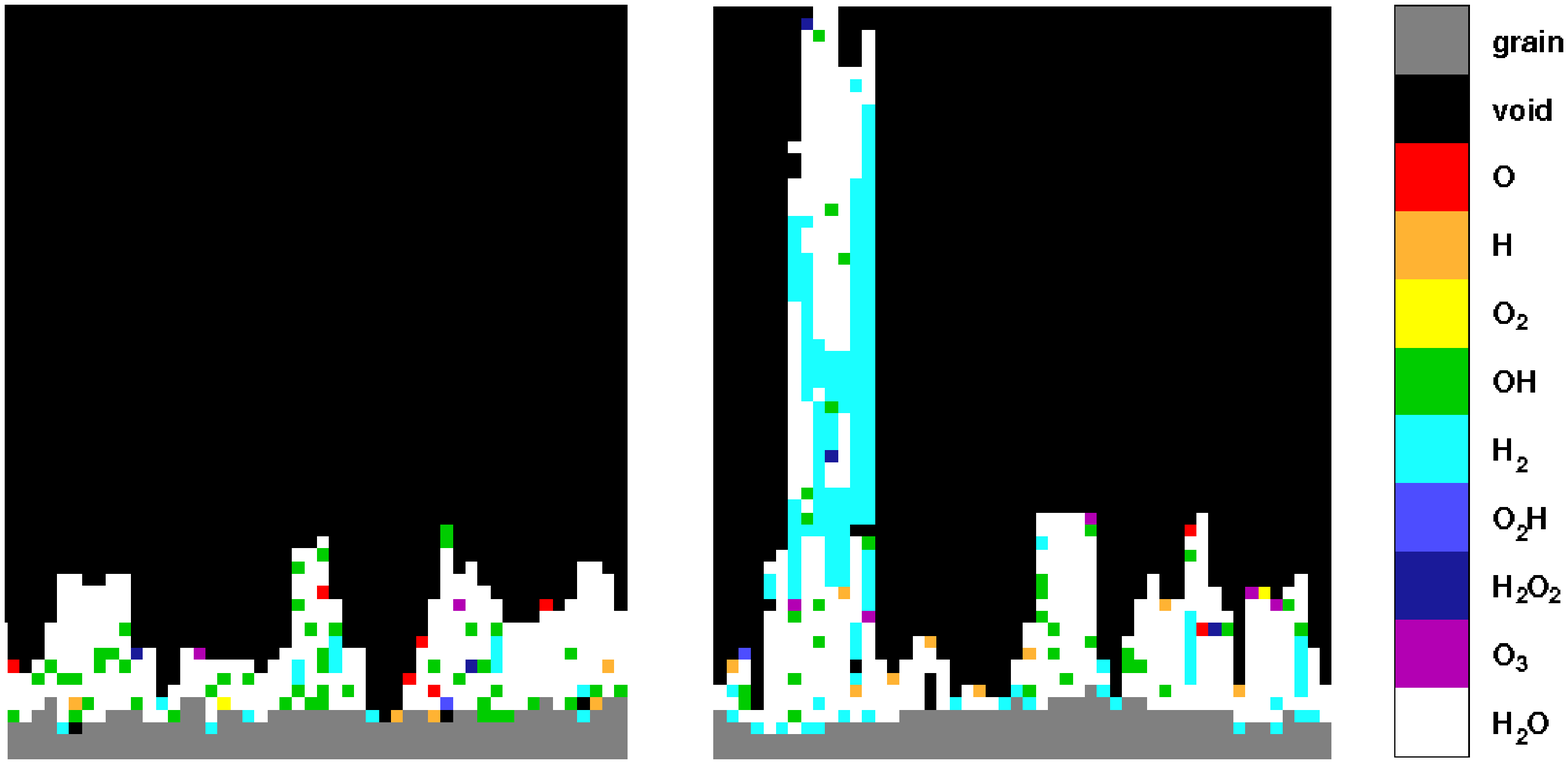}
\caption{Vertical cross sections of the ice mantles after $1 \times 10^5$ yr for two different dense-cloud physical conditions. Left panel (E2): $A_V=5$, $T_{gr} =12$ K, $T_{gas}=20$ K and $n_{\rm H} = 5 \times 10^3$ cm$^{-3}$. Right panel (F2):   $A_V=10$, $T_{gr} =10$ K, $T_{gas}=10$ K and $n_{\rm H} = 2 \times 10^4$ cm$^{-3}$. The molecules are color-coded.}
\label{imagesDense}
\end{figure*}

\begin{figure*}
\plotone{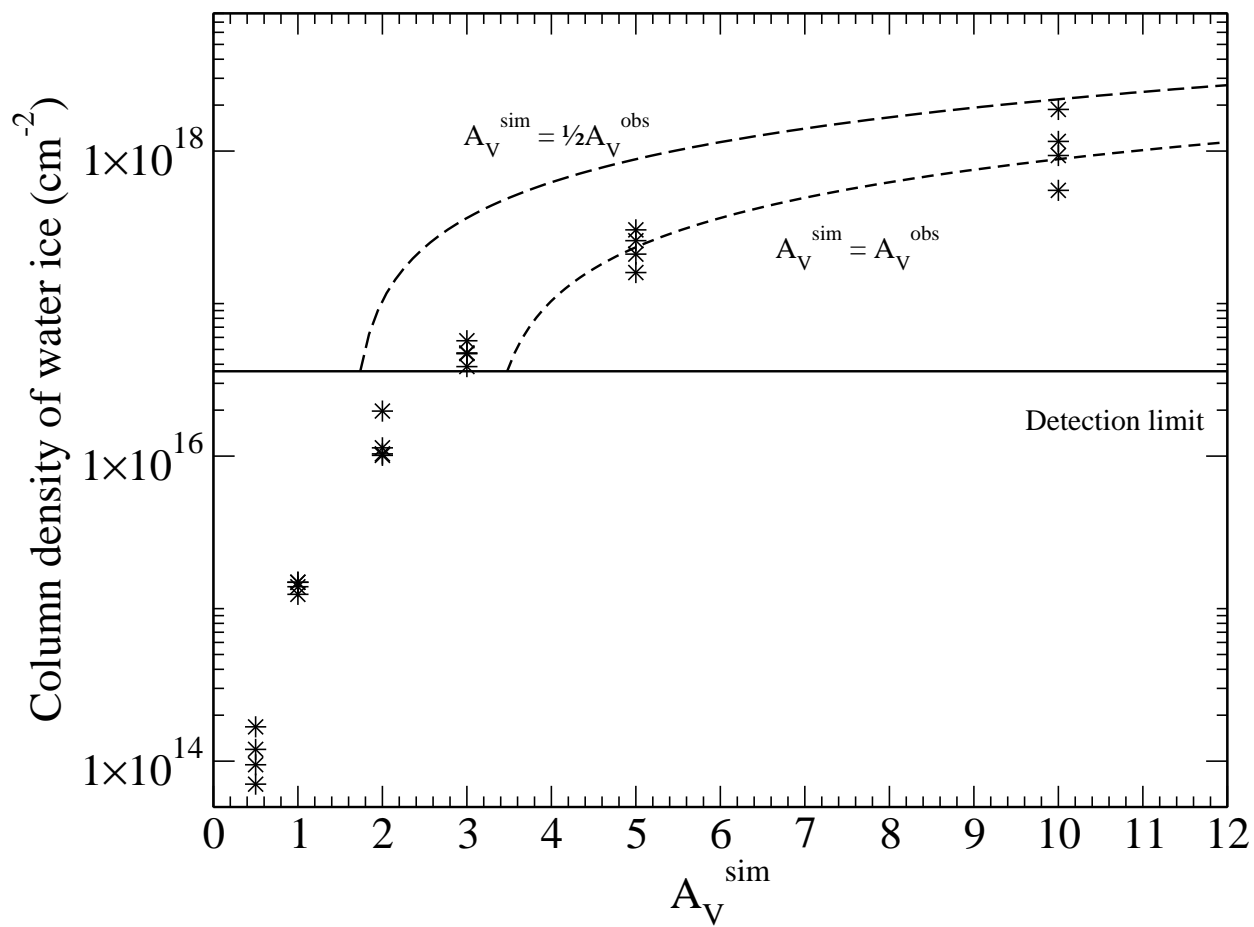}
\caption{Column density of water ice at the end of the simulations ($3 \times 10^5$ yr for $A_V^{sim}=0.5, \dots, 3$ and $10^5$ yr for $A_V^{sim}=5$ and 10) versus the extinction used in our simulations for all physical conditions given in Figs.~\ref{res}-\ref{des}.}  
\label{IcevsAv}
\end{figure*}

\begin{figure*}
\includegraphics[width=0.8\textwidth]{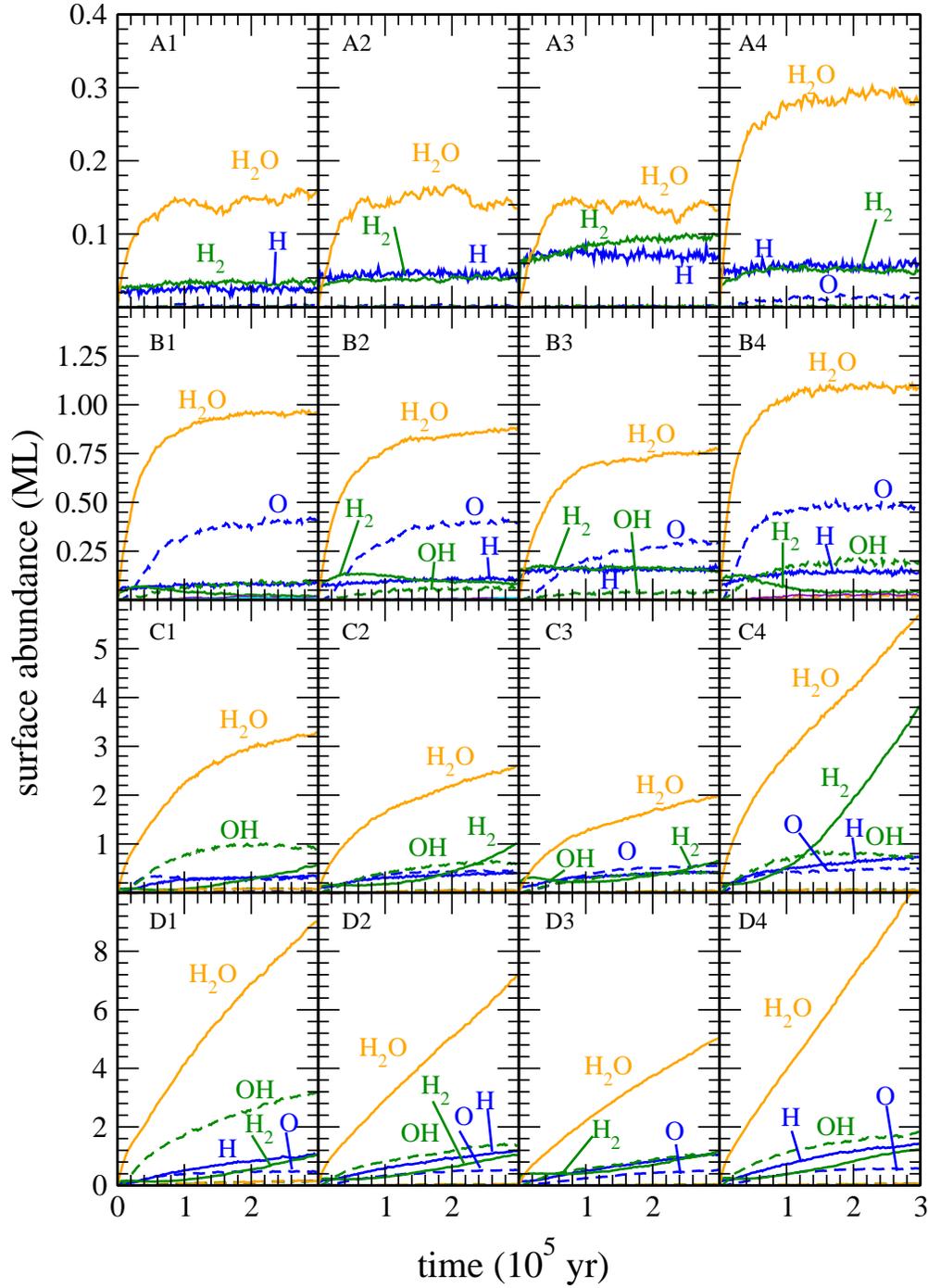}
\caption{The surface abundance with a high hopping barrier-to-desorption energy of 0.78 as a function of time for different conditions in a diffuse or translucent molecular cloud. }
\label{desA}
\end{figure*}

\begin{figure*}
\plotone{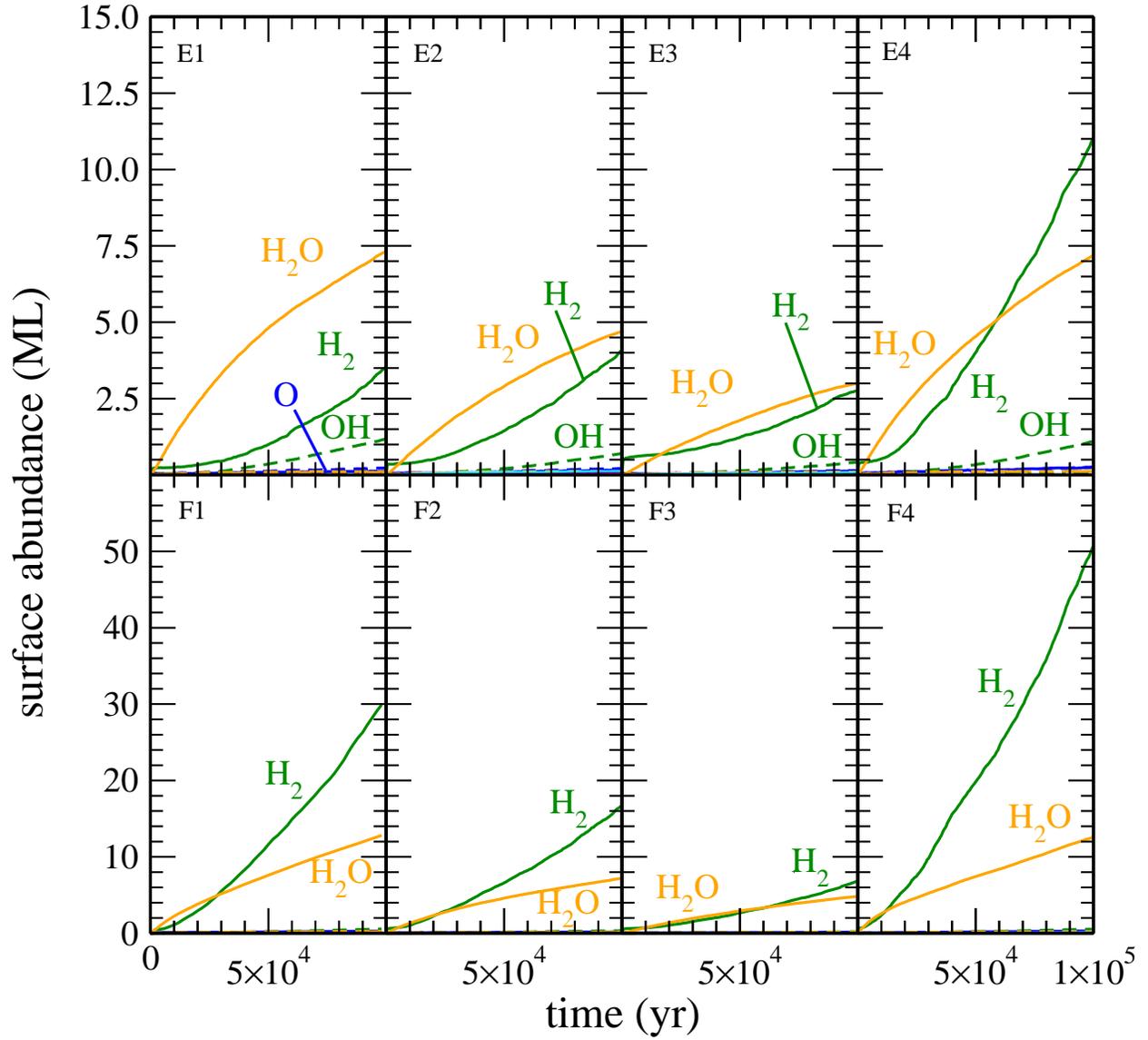}
\caption{The surface abundance with a high hopping barrier-to-desorption energy of 0.78 as a function of time for different conditions in a dense molecular cloud. See Table \ref{phys} for physical conditions in each panel.}
\label{resA}
\end{figure*}

\begin{figure*}
\plotone{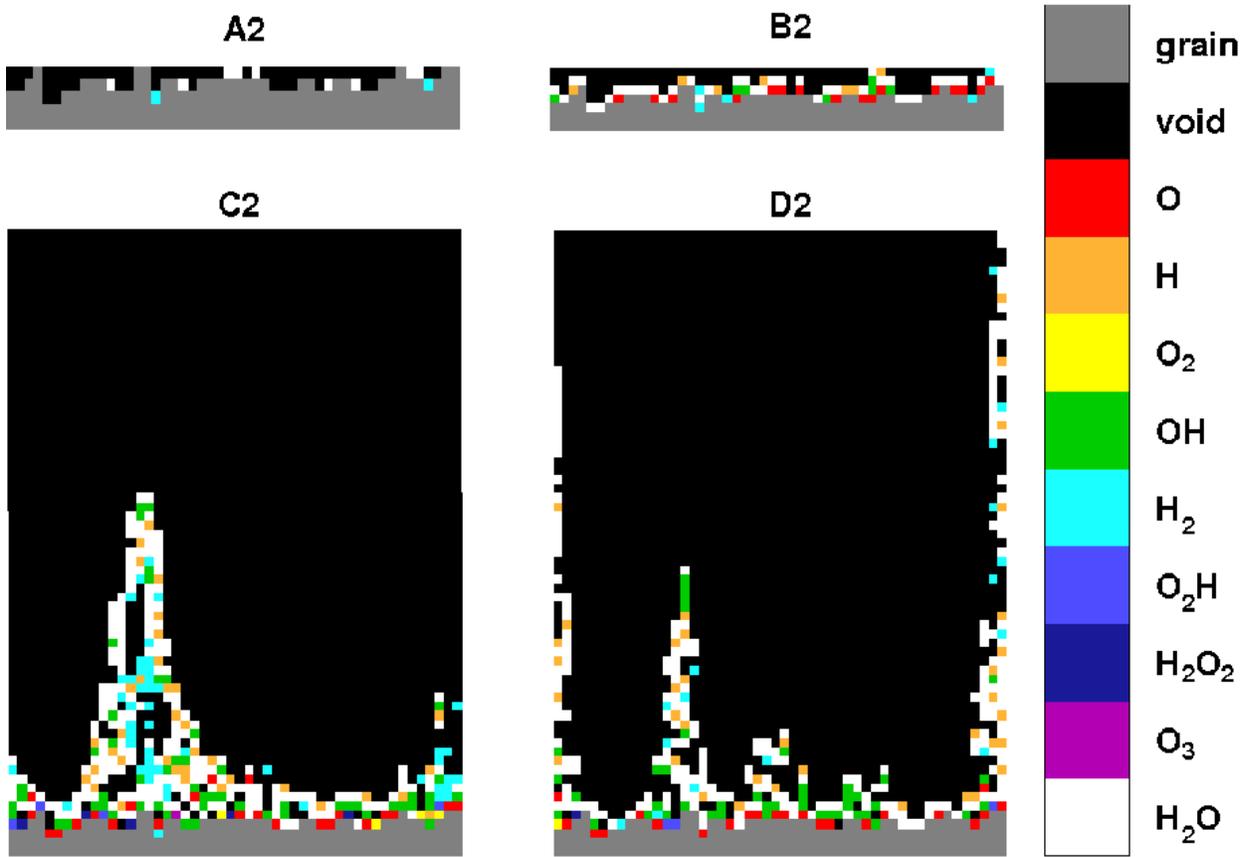}
\caption{Vertical cross sections of  ice mantles after $3 \times 10^5$ yr for four different physical conditions in diffuse and translucent clouds (A2, B2, C2, D2; Table \ref{phys}) obtained with a high hopping barrier-to-desorption energy of 0.78. The molecules are color-coded.}
\label{imagesA}
\end{figure*}

\begin{figure*}
\plotone{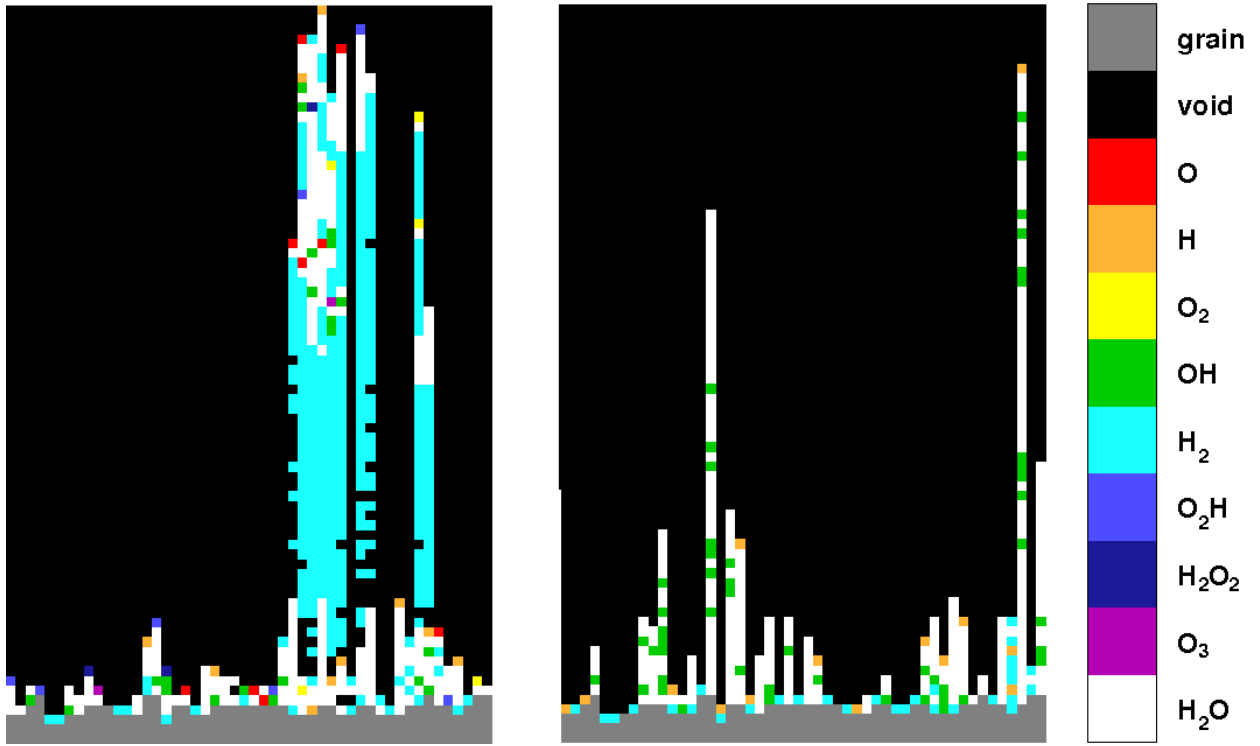}
\caption{Vertical cross sections of ice mantles after $1 \times 10^5$ yr for two different dense-cloud physical conditions obtained with a high hopping barrier-to-desorption energy of 0.78.  Left panel (E2): $A_V=5$, $T_{gr} =12$ K, $T_{gas}=20$ K and $n_{\rm H} = 5 \times 10^3$ cm$^{-3}$. Right panel (F2):   $A_V=10$, $T_{gr} =10$ K, $T_{gas}=10$ K and $n_{\rm H} = 2 \times 10^4$ cm$^{-3}$.  The molecules are color coded.}
\label{imagesDenseA}
\end{figure*}

\begin{figure*}
\plotone{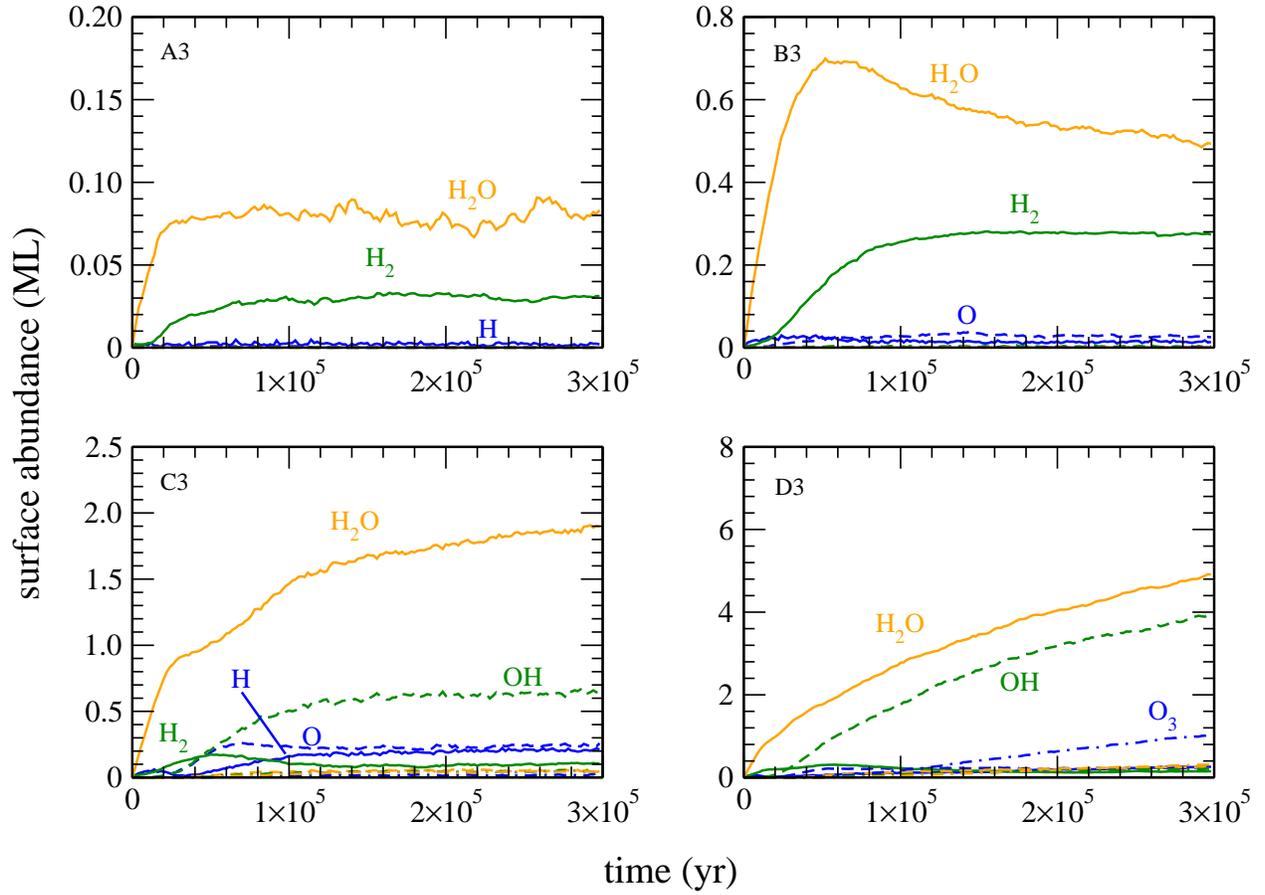}
\caption{The surface abundance as a function of time starting with a bare grain with a smooth surface structure.}
\label{islands}
\end{figure*}

\clearpage

\begin{deluxetable}{lclclllr}
\tablecolumns{3}
\tablewidth{0pc}
\tabletypesize{\scriptsize}
\tablecaption{SURFACE REACTIONS
\label{sur_re}}
\tablehead{\multicolumn{8}{c}{Surface reactions}\\
\multicolumn{6}{c}{Reaction} & \colhead{$\beta$} & \colhead{$E_a$ (K)}}
\startdata
H    &+& H          &$\rightarrow$& H$_2$      & & 0.991\tablenotemark{a} &  0      \\
H    &+& O          &$\rightarrow$& OH         & & 0.991\tablenotemark{a} &  0      \\
H    &+& OH         &$\rightarrow$& H$_2$O     & & 0.991\tablenotemark{a} &  0      \\
O    &+& O          &$\rightarrow$& O$_2$      & & 0.991\tablenotemark{a} &  0      \\
H    &+& O$_2$      &$\rightarrow$& O$_2$H     & & 0.991\tablenotemark{a} &  1200\tablenotemark{b}   \\
H    &+& O$_2$H     &$\rightarrow$& H$_2$O$_2$ & & 0.991\tablenotemark{a} &  0      \\
H    &+& H$_2$O$_2$ &$\rightarrow$& H$_2$O &+ OH & 1         & 1400\tablenotemark{c}   \\
H    &+& O$_3$      &$\rightarrow$& O$_2$  &+ OH & 1         & 450\tablenotemark{d}     \\
H$_2$&+& OH         &$\rightarrow$& H$_2$O &+ H  & 1         & 2600\tablenotemark{e}   \\
O    &+& O$_2$      &$\rightarrow$& O$_3$      & & 0.991\tablenotemark{a} &  0      \\
\enddata
\tablenotetext{a}{Based on photodissociation of water results by \cite{Kroes:2006}.}

\tablenotetext{b}{\cite{Melius:1979}}
\tablenotetext{c}{\cite{Klemm:1975}}
\tablenotetext{d}{\cite{Lee:1978}}
\tablenotetext{e}{\cite{Schiff:1973}}

\end{deluxetable}

\clearpage

\begin{deluxetable}{lclclllr}
\tablecolumns{4}
\tablewidth{0pc}
\tabletypesize{\scriptsize}
\tablecaption{PHOTODISSOCIATION REACTIONS
\label{dis_re}}
\tablehead{\multicolumn{5}{l}{Reaction\tablenotemark{a}} & \colhead{$\alpha_{photo}$ (s$^{-1}$)} & \colhead{$\gamma_{photo}$} & \colhead{$\alpha_{cr}$}}
\startdata
OH	   &$\rightarrow$& O  &+& H	 & 1.68(-10) &  1.66 & 1.02(3)\\
H$_2$O     &$\rightarrow$& H  &+& OH	 & 3.28(-10) &  1.63 & 1.94(3)\\
O$_2$	   &$\rightarrow$& O  &+& O	 & 3.30(-10) & 1.4   & 1.50(3)\\
O$_2$H     &$\rightarrow$& O  &+& OH	 & 	     &       & 1.50(3)\\
O$_2$H     &$\rightarrow$& H  &+& O$_2$  &  	     &       & 1.50(3)\\
H$_2$O$_2$ &$\rightarrow$& OH &+& OH	 &  	     &       & 3.00(3)\\
O$_3$	   &$\rightarrow$& O$_2$&+&  	 & 1.9(-9)   & 1.85\tablenotemark{b}  &  \\
\enddata
\tablenotetext{a}{gas-phase photodissociation rates and products are used}
\tablenotetext{b}{\cite{Dishoeck:2006}}
\end{deluxetable}

\clearpage

\begin{deluxetable}{lrcrrrrrrrrr}
\tablecolumns{8}
\tablewidth{0pc}
\tabletypesize{\scriptsize}
\tablecaption{DESORPTION ENERGIES IN KELVINS
\label{E_eva}}
\tablehead{              & \multicolumn{11}{c}{Substrate} \\ \cline{2-12}
Absorbate  & carbon &&  H & O   & OH  & H$_2$& O$_2$& H$_2$O& O$_3$& O$_2$H& H$_2$O$_2$}
\startdata
H          &  658 & & ... & ... & ... &  45  &   45 &  650  &  45  & ...   &  45\\ 
O          &  800 & & ... & ... &  55 &  55  &   55 &  800  &  55  &  55   &  55\\ 
OH         & 1360 & & ... & 240 & 240 & 240  &  240 & 3500  & 240  & 240   & 240\\ 
H$_2$      &  542 & &  30 &  30 &  30 &  23  &   30 &  440  &  30  &  30   &  30\\ 
O$_2$      & 1440 & &  69 &  69 &  69 &  69  &   69 & 1000  &  69  &  69   & 109\\ 
H$_2$O     & 2000 & & 390 & 390 & 390 & 390  &  390 & 5640  & 390  & 390   & 390\\ 
O$_3$      & 2240 & & 120 & 120 & 120 & 120  &  120 & 1800  & 120  & 120   & 120\\ 
O$_2$H     & 2160 & & ... & 300 & 300 & 300  &  300 & 4300  & 300  & 300   & 300\\ 
H$_2$O$_2$  & 2818 & & 340 & 340 & 340 & 340  &  340 & 4950  & 340  & 340   & 340\\ 
\enddata
\tablecomments{See discussion in text}
\end{deluxetable}

\clearpage

\begin{deluxetable}{cccccc}
\tablecolumns{8}
\tablewidth{0pc}
\tabletypesize{\scriptsize}
\tablecaption{PHYSICAL CONDITIONS OF CHOSEN MODELS\label{phys}}
\tablehead{\colhead{Panel}   & \colhead{$A_V$} & \colhead{$T_{grain}$ (K)} & \colhead{$T_{gas}$ (K)} & \colhead{$n_{\rm H}$ (cm$^{-3}$} )    & \colhead{Form of H}}
\startdata
A1 & 0.5   & 20 	 & 100       & 1.0(2)	  &  H\\
A2 & 0.5   & 18 	 & 80	     & 1.0(2)	  &  H\\
A3 & 0.5   & 16 	 & 60	     & 1.0(2)	  &  H\\
A4 & 0.5   & 18 	 & 80	     & 5.0(2)	  &  H\\
B1 & 1.0     & 18 	 & 80	     & 2.5(2)	  &  H\\
B2 & 1.0     & 16 	 & 60	     & 2.5(2)	  &  H\\
B3 & 1.0     & 14 	 & 40	     & 2.5(2)	  &  H\\
B4 & 1.0     & 16 	 & 60	     & 5.0(2)	  &  H\\
C1 & 2.0     & 17 	 & 70	     & 5.0(2)	  &  H\\
C2 & 2.0     & 15 	 & 50	     & 5.0(2)	  &  H\\
C3 & 2.0     & 13 	 & 30	     & 5.0(2)	  &  H\\
C4 & 2.0     & 15 	 & 50	     & 1.0(3)	  &  H\\
D1 & 3.0     & 16 	 & 60	     & 1.0(3)	  &  H\\
D2 & 3.0     & 14 	 & 40	     & 1.0(3)	  &  H\\
D3 & 3.0     & 12 	 & 20	     & 1.0(3)	  &  H\\
D4 & 3.0     & 14 	 & 40	     & 1.5(3)	  &  H\\
\hline
E1 & 5.0     & 14 	 & 40	    & 5.0(3)	 &  H$_2$\\
E2 & 5.0     & 12 	 & 20	    & 5.0(3)	 &  H$_2$\\
E3 & 5.0     & 10 	 & 10	    & 5.0(3)	 &  H$_2$\\
E4 & 5.0     & 12 	 & 20	    & 1.0(4)	 &  H$_2$\\
F1 & 10.0    & 12 	 & 20	    & 2.0(3)	 &  H$_2$\\
F2 & 10.0    & 10 	 & 10	    & 2.0(4)	 &  H$_2$\\
F3 & 10.0    & 10 	 & 10	    & 1.0(4)	 &  H$_2$\\
F4 & 10.0    & 10 	 & 10	    & 5.0(4)	 &  H$_2$\\
\enddata
\tablecomments{The notation a(b) implies $a\times10^{b}$.}
\end{deluxetable}

\clearpage

\begin{deluxetable}{ccccccccccccc}
\tablecolumns{9}
\tablewidth{0pc}
\tabletypesize{\scriptsize}
\tablecaption{H$_2$ RECOMBINATION EFFICIENCY
\label{etha}}
\tablehead{ \colhead{} & \colhead{} & \multicolumn{2}{c}{1} & \colhead{} & \multicolumn{2}{c}{2} & \colhead{} & \multicolumn{2}{c}{3} & \colhead{} & \multicolumn{2}{c}{4}}
\startdata
A && 0.05 & 0.02 && 0.04 & 0.17 && 0.22 & 0.57 && 0.05 & 0.11 \\
B && 0.06 & 0.28 && 0.15 & 0.59 && 0.24 & 0.72 && 0.17 & 0.60 \\
C && 0.24 & 0.54 && 0.29 & 0.65 && 0.48 & 0.71 && 0.22 & 0.68 \\
D && 0.11 & 0.61 && 0.43 & 0.74 && 0.52 & 0.71 && 0.54 & 0.73 \\
\enddata
\tablecomments{The left columns are results in the presence of an oxygen flux, the right in its absence.}
\end{deluxetable}

\clearpage

\begin{deluxetable}{ccrrrcrrrcrrrcrrr}
\tablecolumns{9}
\tablewidth{0pc}
\tabletypesize{\scriptsize}
\tablecaption{CONTRIBUTIONS OF THREE REACTIONS TO WATER FORMATION
\label{reactions}}
\tablehead{ \colhead{} & \colhead{} & \multicolumn{3}{c}{1} & \colhead{} & \multicolumn{3}{c}{2} & \colhead{} & \multicolumn{3}{c}{3} & \colhead{} & \multicolumn{3}{c}{4}}
\startdata
A && 99.9 &  0.0 &  0.1 && 99.6 &  0.0 &  0.4 && 95.9 &  0.0 &  4.2 && 99.5 &  0.0 &  0.5 \\
B && 99.5 &  0.0 &  0.5 && 98.3 &  0.0 &  1.7 && 96.4 &  0.0 &  3.6 && 98.5 &  0.1 &  1.4 \\
C && 98.4 &  0.5 &  1.1 && 97.1 &  0.8 &  2.0 && 93.1 &  1.0 &  5.9 && 95.9 &  1.3 &  2.9 \\
D && 97.6 &  1.4 &  1.0 && 93.4 &  1.4 &  5.2 && 90.5 &  1.6 &  7.9 && 93.2 &  0.9 &  5.9\\
\hline
E && 14.3 & 21.0 & 64.7 && 11.6 & 19.3 & 69.1 &&  7.0 & 16.7 & 76.3 &&  9.7 & 18.3 & 72.0 \\
F && 10.1 & 18.7 & 71.2 &&  6.5 & 16.7 & 76.8 &&  6.2 & 16.5 & 77.2 &&  6.7 & 16.2 & 77.1\\
\enddata
\tablecomments{From left to right: H + OH $\rightarrow$ H$_2$O, H + H$_2$O$_2$ $\rightarrow$ H$_2$O + OH, and H$_2$ + OH $\rightarrow$ H$_2$O + H.}
\end{deluxetable}

\clearpage

\begin{deluxetable}{ccccccccccccc}
\tablecolumns{9}
\tablewidth{0pc}
\tabletypesize{\scriptsize}
\tablecaption{H$_2$ RECOMBINATION EFFICIENCY WITH SLOW DIFFUSION
\label{ethaA}}
\tablehead{ \colhead{} & \colhead{} & \multicolumn{2}{c}{1} & \colhead{} & \multicolumn{2}{c}{2} & \colhead{} & \multicolumn{2}{c}{3} & \colhead{} & \multicolumn{2}{c}{4}}
\startdata
A && 0.47 & 0.56 && 0.58 & 0.72 && 0.61 & 0.68 && 0.59 & 0.69 \\
B && 0.67 & 0.74 && 0.64 & 0.67 && 0.68 & 0.71 && 0.67 & 0.68 \\
C && 0.40 & 0.71 && 0.43 & 0.74 && 0.49 & 0.65 && 0.37 & 0.75 \\
D && 0.30 & 0.70 && 0.34 & 0.68 && 0.32 & 0.69 && 0.31 & 0.67 \\
\enddata
\tablecomments{The left columns are results in the presence of an oxygen flux, the right in its absence.The calculations are done with a ratio of 0.78 between hopping barrier and desorption energy.}
\end{deluxetable}

\end{document}